\documentstyle[11pt, epsf]{article}

\newcommand{\be}{\begin{equation}}
\newcommand{\ee}{\end{equation}}
\newcommand{\bea}{\begin{eqnarray}}
\newcommand{\eea}{\end{eqnarray}}
\def\mgh{m_{\rm gh}}
\def\mW{m_W}
\def\mWq{m_W^2}
\def\mWf{m_W^4}
\def\mGs{m_{\rm Gs}}
\def\mGsq{m_{\rm Gs}^2}
\def\mGsf{m_{\rm Gs}^4}
\def\mH{m_{\rm H}}
\def\mHq{m_{\rm H}^2}
\def\mHf{m_{\rm H}^4}

\begin{document}

\begin{titlepage}
\begin{flushright}
HD--THEP--95--1\\
hep-ph/9501317
\end{flushright}
\vspace{1.5cm}
\begin{center}
\mbox{\bf\LARGE The High-Temperature Two-Loop Effective Potential of the}\\
\vspace{.2cm}
{\bf\LARGE  Electroweak Theory
in a General 't~Hooft Background Gauge}\\
\vspace{1cm}
{\bf
Jochen Kripfganz\footnote{supported by Deutsche
Forschungsgemeinschaft}\\
\vspace{.3cm}
Andreas Laser\footnote{supported by Landesgraduiertenf\"orderung
Baden-W\"urttemberg\\[1ex]
e-mail addresses:\\[0.5ex]
\begin{tabular}{rl}
J.~Kripfganz & dj8@vm.urz.uni-heidelberg.de\\
A.~Laser & t82@ix.urz.uni-heidelberg.de\\
M.G.~Schmidt & k22@vm.urz.uni-heidelberg.de
\end{tabular}}\\
\vspace{.3cm}
Michael G.~Schmidt\\
}
\vspace{1cm}
Institut  f\"ur Theoretische Physik\\
Universit\"at Heidelberg\\
Philosophenweg 16\\
D-69120 Heidelberg, FRG\\
\vspace{1.5cm}
{\bf Abstract}\\
\end{center}
We calculate the high-temperature two-loop effective potential
using a general 't~Hooft background gauge. The dependence
on the gauge-fixing parameter $\xi$ is investigated.
The effective coupling constant at the critical temperature $g_3(T_c)^2$
is decreased considerably  compared to the one-loop
result, independent of $\xi$.
\end{titlepage}

There are strong indications that the electroweak standard theory predicts
a first order phase transition at the electroweak scale
\cite{KirzhnitsLi}-\cite{FarakosEA2}.
Generating the baryon asymmetry of the universe at the electroweak phase
transition is an exciting possibility. A better understanding of this phase
transition
is required, however, in order to clarify whether this is indeed the case.
The electroweak phase
transition cannot be treated completely by perturbative techniques. Problems
are
caused by infrared singularities of the symmetric phase, requiring the
summation
of infinite sets of diagrams. Lattice simulations take care of this
automatically, and
therefore are indispensable tools for the investigation of the electroweak
phase transition \cite{BunkEA}-\cite{FarakosEA2}.
However, they are not well suited for the study of important
physical quantities like the sphaleron transition rate, or the rate of critical
bubble formation.
These quantities have been
studied in quasiclassical approximation to one-loop order.
Two-loop calculations should
help to control to
what extend the corresponding results are reliable,
though of course genuine nonperturbative contributions to the
potential in the infrared have to be taken into account differently.

Both sphalerons and critical bubbles are static field configurations, i.e. they
do
not depend on the imaginary time variable $\tau$. Therefore, it becomes useful
to integrate out the non-static Matsubara frequencies first, to some order in
the loop expansion. This perturbative expansion should be reliable because
the nonstatic Matsubara frequencies become heavy at high temperature. The
longitudinal component of the gauge field $A_0$
develops a Debye-mass proportional to $gT$ and may be integrated out
as well \cite{JakovacEA}.

The resulting  three-dimensional effective theory is of course non-local.
Usually higher derivative terms are neglected in the spirit of the
high temperature expansion.
One also neglects the Weinberg mixing and
considers the action of the three-dimensional SU(2)-Higgs model:
\be\label{Sht}
S_{\rm ht} = \frac{1}{g_3(T)^2}  \int d^3x \left[ \frac{1}{4} F_{ij}^a F_{ij}^a
 + (D_i\Phi)^\dagger (D_i\Phi) +
 V_{\rm ht}(\Phi^\dagger\Phi) \right] \quad,
\ee
where we have introduced dimensionless coordinates and fields
\be \label{rescale}
\vec{x} \rightarrow \frac{\vec{x}}{gv}, \qquad   \Phi \rightarrow v\Phi,
 \qquad   A \rightarrow v A \quad.
\ee
The scale $v$ is left open for the moment.

The effective 3-dimensional gauge coupling is
defined as
\be
g_3(T)^2 = \frac{g T}{v} \quad.
\ee
The gauge coupling $g$ has been scaled out of the covariant derivative
and the field strength tensor.
The high temperature effective potential is
\begin{equation}
V_{\rm ht}(\Phi^\dagger \Phi) = \lambda\left((\Phi^\dagger \Phi)^2 -
v_0^2 \Phi^\dagger \Phi\right) \quad.
\end{equation}
For compactness of notation we use a rescaled ``$\lambda$'' and ``$v_0^2$''.
They are temperature dependent constants which correspond as
$\lambda \cong \lambda_T/g^2$ and  $v_0^2 \cong (v_0(T)/v)^2$ to the
one-loop quantities used in reference \cite{KripfganzEA},
respectively as $\lambda \cong \bar{\lambda}_3/g_3^2 $ and
$\lambda v_0^2 \cong -\bar{m}_3^2/v^2$ to the two-loop quantities
used in \cite{FarakosEA}.

We divide the fields into a background and fluctuations
\begin{eqnarray}
\Phi \rightarrow \hat{\Phi} + g_3 \tilde{\Phi}, &\qquad
\hat{\Phi} = \frac{1}{\sqrt 2} \, \varphi \vec{e}, &\qquad
\tilde{\Phi} = \frac{1}{\sqrt 2}\,(\sigma {\bf 1} + i \pi^a \tau^a) \vec{e} \\
A \rightarrow \hat{A}_i^a + g_3 \tilde{A}_i^a, &\qquad
\hat{A} = 0, &\qquad \tilde{A}_i^a = a_i^a \quad.
\end{eqnarray}
(The $\tau^a$ are the Pauli matrices.) In order to describe critical
bubbles responsible for the onset of the
electroweak phase transition it is sufficient to work with only one
nonvanishing background component $\varphi$ of $\Phi$ in an arbitrary but
constant direction $\vec{e}$. We consider this type of
background only.

Integrating out the fluctuating fields in the loop expansion
generates an effective
action to be used to find the saddle point solutions corresponding to
sphalerons and critical bubbles.  To higher loop order this expansion will
break down for small values of the Higgs field  $\varphi$.
Near the broken minimum,
however, the loop expansion is expected to work quite well.  Whether the
saddle point actions can be estimated reliably depends on how important
different regions in field space are for the corresponding solutions.  This
question deserves further studies.

In praxi it is
not possible to calculate the full effective action, but one has to expand
it in some way, cut off the expansion and calculate the coefficient
in powers of derivatives of the field. This expansion must of course break
down at small values of the field $\varphi$, because the derivative
operator $\partial$ has the mass dimension 1,
which must be compensated by powers of $\varphi^{-1}$.
Indeed calculating the contribution of a higher term in the
derivative expansion to the effective action of a quasiclassical
configuration one finds a divergent result, except for the potential
and the $\partial_i \varphi \partial_i \varphi$ term.
However for the latter kinetic term one finds a $Z$-factor \cite{KripfganzEA}
which in one-loop order is very strongly gauge dependent and therefore
even this term, if considered separately, is rather unphysical.

What seems to be needed is some less gauge dependent (nonlocal)
combination of kinetic terms. As we will demonstrate, the effective potential
is less gauge dependent. In a strict expansion in $g_3^2$ its
extrema are
completely gauge-independent. It will play an  essential role
in the case of inhomogeneous field configurations as well.

In order to integrate out the fluctuating fields one has to fix the gauge.
We choose as gauge-fixing condition the 't~Hooft background gauge
\begin{eqnarray}\label{gauge-fixing}
{\cal F}^a \;=\; D(\hat{A})_i \tilde{A}_i^a +
\frac{i}{2} \xi (\hat{\Phi}^\dagger
\tau^a \tilde{\Phi} - \tilde{\Phi}^\dagger \tau^a \hat{\Phi}) =
\partial_i a_i^a - \frac{1}{2} \xi \varphi \pi^a \quad.
\end{eqnarray}
The resulting effective action
$\Gamma[\varphi]$ depends in general on the gauge-fixing.
Physical quantities should, on the other hand, be independent of it.
This is due to the fact that they are described by extrema of
$\Gamma[\varphi]$. Kobes et.al.~\cite{KobesEA} showed, that the gauge-fixing
dependence
of the effective action can be written as
\be \label{gfindep}
\delta \Gamma[\varphi] \;=\;
\frac{\delta \Gamma[\varphi]}{\delta\varphi}
\,\delta X[\varphi] \quad,
\ee
where $\varphi$ represents all kinds of fields and $\delta X[\varphi]$
is a functional of the fields which can  be calculated
from the gauge-fixing condition and from the generators of the gauge
transformation. One reads of immediately that the value of the effective
action is gauge-fixing independent for solutions of the equations of motion,
as it should.

An often raised objection against the 't~Hooft background gauges
is (see e.g.~\cite{Arnold}) that the field $\tilde{\varphi}$
used in the gauge-fixing should be of another type than the background field
and should therefore not be varied in calculating the equation of
motion
\be
\left( \frac{\delta \Gamma[\varphi,\tilde{\varphi}]}{\delta\varphi}
\right)_{\tilde{\varphi}=\varphi}
\;\neq\;  \frac{\delta \Gamma[\varphi,\varphi]}{\delta\varphi} \quad.
\ee
Nevertheless equation (\ref{gfindep}) holds even
for these class of gauges with
$\Gamma[\varphi] = \Gamma[\varphi,\tilde{\varphi}\!=\!\varphi]$
as is explicitly shown in reference \cite{KobesEA}.
Similar statements, although less general, have been verified
long time ago \cite{Nielsen,FukudaEA}.

As a consistency check, we shall explicitly demonstrate the
gauge-fixing independence (i.e.~the $\xi$-independence) of
the value of the effective potential at its extrema.
In order to achieve this one has to be careful to work consistently
to a given order of $g_3^2$.

Expanding in terms of the fluctuating fields we obtain the action
\begin{eqnarray}\label{Sfluct}
S_{\rm ht} &+& \int d^3x \left( \frac{1}{2\xi} {\cal F}^a {\cal F}^a +
	{\cal L}_{FP} \right)
\quad \rightarrow  \quad \int d^3x \bigg\{
 \\
\frac{1}{g_3^2} &\bigg( &
	\frac{1}{2}\partial_i\varphi \partial_i\varphi
	+ V\Big(\frac{1}{2} \varphi^2 \Big)
\quad \bigg) \; + \nonumber \\
\frac{1}{g_3} &\bigg(&
	\partial_i\varphi\partial_i\sigma
	+ \lambda \left( \varphi^2 - v_0^2 \right) \sigma \varphi
\quad \bigg) \; + \nonumber \\
&\bigg(&
	\frac{1}{2} \left(\partial_i a_j^a\right)^2
	+ \frac{1}{2\xi} \left(1-\xi\right) \left(\partial_i a_i^a\right)^2
	+ \frac{1}{8} \varphi^2 a_i^a a_i^a
	+ \frac{1}{2} \partial_i\sigma \partial_i\sigma
	+ \frac{1}{2} \lambda \left(3 \varphi^2- v_0^2\right) \sigma^2
	\nonumber \\ &
	+ & \left(\partial_i \pi^a\right)^2
	+ \frac{1}{2} \left( \frac{1}{4} \xi \varphi^2
	+ \lambda\left(\varphi^2-v_0^2\right) \right) \pi^a \pi^a
	- c^{a*}\partial^2 c^a
	+ \frac{1}{4} \xi \varphi^2 c^{a*}c^a
	+ a_i^a \pi^a \partial_i\varphi
\;\bigg)\; + \nonumber \\
g_3 &\bigg(&
	\frac{1}{2} \epsilon^{abc} a_i^a a_j^b
		\left(\partial_i a_j^c - \partial_j a_i^c \right)
	+ \frac{1}{2} \epsilon^{abc} a_i^a \pi^b \partial_i \pi^c
	- \epsilon^{abc} \partial_i c^{a*} c^b a_i^c
	- \frac{1}{4} \xi \epsilon^{abc} \varphi \pi^a c^{b*} c^c
	\nonumber \\ &&
	+ \lambda \left( \sigma^2 + \pi^a \pi^a \right) \sigma \varphi
	+ \frac{1}{2} a_i^a \left( \partial_i \sigma \pi^a -
		\sigma \partial_i \pi^a \right)
	+ \frac{1}{4} \varphi \sigma a_i^a a_i^a
	+ \frac{1}{4} \xi \varphi \sigma c^{a*} c^a
\quad\bigg) \;+  \nonumber \\
g_3^2 &\bigg( &
	\frac{1}{4} \epsilon^{abc} \epsilon^{ade} a_i^b a_j^c a_i^d a_j^e
	+ \frac{1}{8} a_i^a a_i^a \pi^b \pi^b
	+ \frac{1}{4} \lambda \left( \sigma^2 + \pi^a \pi^a \right)^2
	+ \frac{1}{8} \left( \sigma^2 + \pi^a \pi^a \right) a_i^b a_i^b
\quad \bigg)  \; \;
\bigg\} \nonumber
\end{eqnarray}

One reads off the propagators
and vertices. The gauge boson $(a_i^a)$ propagator may be written as
\begin{eqnarray}\label{gauge_prop}
D_W(k)_{ij}^{ab} &=& \delta^{ab} \left( \frac{1}{k^2 + m_W^2}
\left(\delta_{ij} + \frac{k_i k_j}{m_W^2} \right)
\;-\; \frac{k_i k_j}{m_W^2}\; \frac{1}{k^2 + \xi m_W^2} \right) \quad.
\end{eqnarray}
The Higgs $(\sigma)$, Goldstone $(\pi^a)$ and ghost $(c^a)$ propagators
are as usual with the masses
\begin{eqnarray}\label{masses}
&m_{\rm H}^2 \;=\; \lambda (3\varphi^2 - v_0^2) \quad,
\qquad &m_{\rm Gs}^2 \;=\;
\lambda (\varphi^2 - v_0^2) + \frac{1}{4} \xi \varphi^2 \quad, \\
&m_W^2 \;=\; \frac{1}{4} \varphi^2 \quad,
\qquad &m_{\rm gh}^2 \;=\; \frac{1}{4} \xi \varphi^2 \quad.
\end{eqnarray}

Note that there are no IR-divergences due to massless Goldstone-bosons
at the broken minimum,  for $\xi\neq 0$.
In the background field formalism the fluctuation appear only
in the inner lines while the external lines consist of background
fields \cite{Abbott}. Only the 1PI-graphs  contribute to the effective
action.

In the following we study the effective potential.
Up to now it has been calculated to two-loop order in
Landau gauge from the four-dimensional theory \cite{ArnoldEs,FodorHe}
and from the high-temperature theory \cite{FarakosEA}.
In a recent work M.~Laine   \cite{Laine} presented a calculation using
a general covariant gauge (${\cal F}^a = \partial_i A_i^a$).

The aim of our work is to calculate the potential in two-loop
order using the 't~Hooft background gauge with an arbitrary gauge-fixing
parameter $\xi$, and to investigate the $\xi$-dependence.

At one-loop one obtains the effective potential
\begin{eqnarray}
V_{\rm eff}^{(1)} \;=\; \lambda (\frac{1}{4} \varphi^4 -
\frac{1}{2} v_0^2 \varphi^2) - \frac{g_3^2}{12 \pi}
( m_{\rm H}^3 + 3 m_{\rm Gs}^3 + (6 + 3 \xi^{3/2}) m_W^3 - 6 m_{\rm gh}^3)
\quad.
\end{eqnarray}

\begin{figure}[t]
\epsfxsize14cm
\epsffile{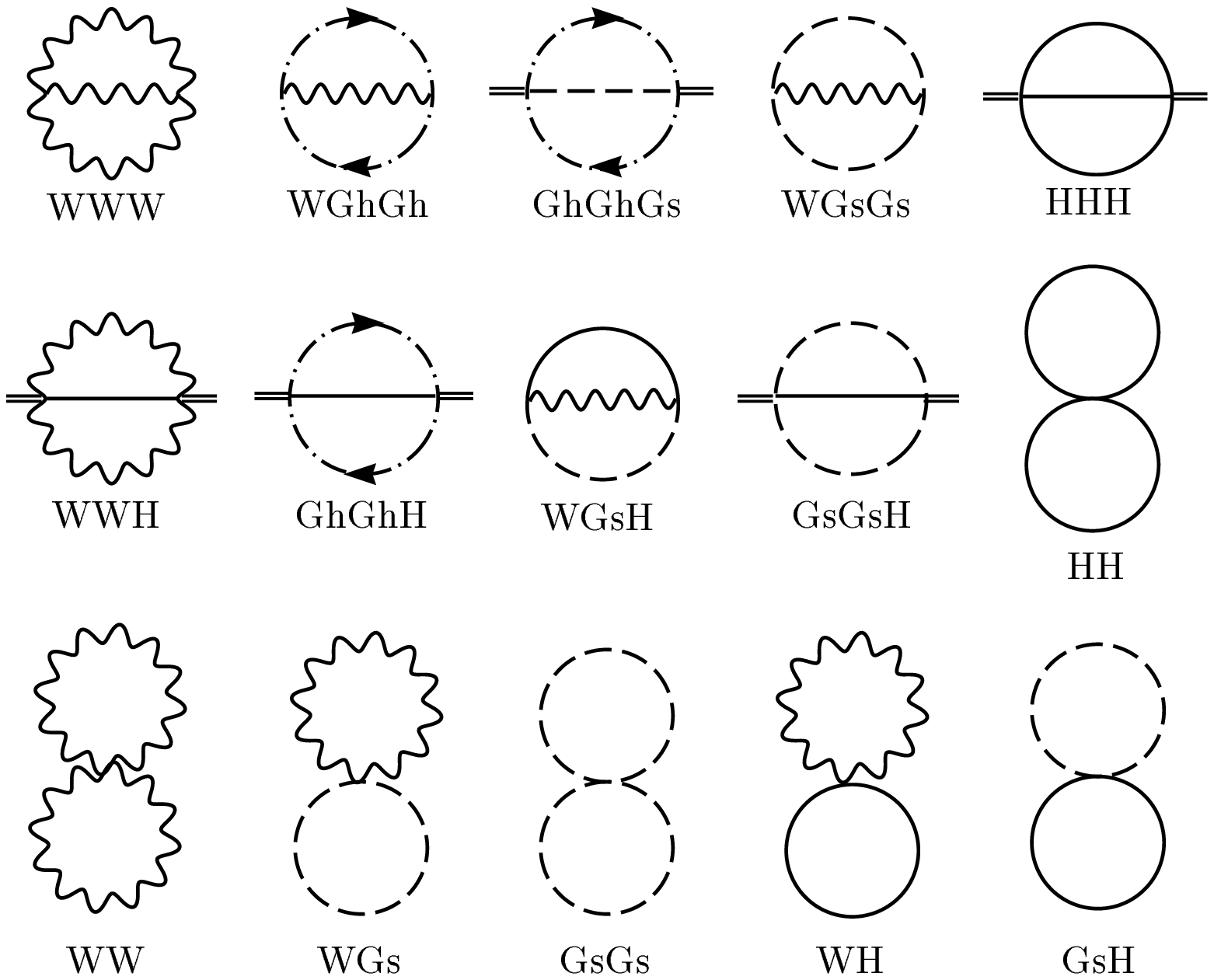}
\begin{tabular}{rll}
&&\\
The lines are: &
\raisebox{-0.05cm}{\epsfxsize1cm \epsffile{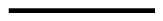}} Higgs ($\sigma$);&
\raisebox{-0.05cm}{\epsfxsize1cm \epsffile{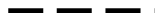}} Goldstone
($\pi^a$);\\ &
\raisebox{-0.1cm}{\epsfxsize1cm \epsffile{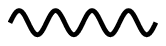}} gauge boson
($a_i^a$);&
\raisebox{-0.2cm}{\epsfxsize1cm \epsffile{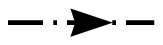}} ghost ($c^a$);\\&
\raisebox{-0.1cm}{\epsfxsize1cm \epsffile{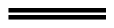}}
background field ($\varphi$) &
\end{tabular}
\caption{The Graphs contributing to the two-loop effective potential.}
\label{FIGgraphs}
\end{figure}

The two-loop potential receives contributions from the graphs shown in
figure \ref{FIGgraphs}.
We use  the $\overline{\rm MS}$-renormalization scheme.
The sunset
graphs consist of integrals with three denominators; the figure-eight
graphs have two denominators. The latter can be calculated easily while the
momenta in the numerators of the sunsets cause some trouble.
We removed them by a procedure which is similar to the one used in reference
\cite{FarakosEA}.
The remaining integrals are of the following form ($d = 3 - 2 \epsilon$)
\begin{eqnarray}
G(m_1,m_2) &=& \mu^{4\epsilon} \int \frac{d^dk}{(2 \pi)^d}
\int \frac{d^dp}{(2 \pi)^d}\;
\frac{1}{k^2 + m_1^2}  \; \frac{1}{p^2 + m_2^2} \quad=\quad
\frac{1}{16 \pi^2}\; m_1 \; m_2  \label{IntG}\\
H(m_1,m_2,m_3) &=&  \mu^{4\epsilon} \int \frac{d^dk}{(2 \pi)^d}
\int \frac{d^dp}{(2 \pi)^d}\;
\frac{1}{k^2 + m_1^2}\; \frac{1}{p^2 + m_2^2}\;
\frac{1}{(k + p)^2+ m_3^2} \label{IntH}\\
&=& \frac{1}{32 \pi^2} \left(\frac{1}{2 \epsilon} -
	\gamma + \ln(4 \pi)\right) +
\frac{1}{16 \pi^2} \left(\frac{1}{2} -
	\ln\left( \frac{m_1 +  m_2 + m_3}{\mu}\right)\right) \label{Hsol}
\end{eqnarray}

The two-loop potential
is a rather long expression and therefore not displayed here.
All the ingredients are given in the appendix.
While single graphs have $\xi$-dependent divergences the overall
divergence is $\xi$-independent
\be
\frac{9}{2} \lambda \varphi^2
- 6 \lambda^2 \varphi^2
- 3 \lambda v_0^2
+ \frac{51}{32} \varphi^2 \quad.
\ee
{}From equation (\ref{Hsol}) one sees that the $\mu$-dependent part of
the potential is proportional to the divergence and therefore
$\xi$-independent. Hence it is possible to treat the $\mu$-dependence
for one value of $\xi$ and the $\xi$-dependence for one value of $\mu$.
The former is usually discussed by means of the renormalization group.
This has been done for the Landau gauge elsewhere \cite{FarakosEA} and
will not be repeated here.
Instead we are setting $\mu$ to be the value of the field in the
broken minimum later on.

As mentioned above the value of the action must be independent of
the gauge-fixing on its extrema. In our case the value of the potential
at its extrema should be $\xi$-independent. In order to expand it
systematically in orders of $g_3^2$ one has to expand the value of the field
at the minimum first
\be
\varphi_{\rm min} \;=\; \varphi_{\rm min}^{(0)}
	\,+\, g_3^2 \varphi_{\rm min}^{(1)} \,+\, g_3^4 \varphi_{\rm min}^{(2)}
	\,+\, {\cal O}(g_3^6) \quad.
\ee
Plugging this into the effective potential one gets
\be\label{Vmin}
V_{\rm min} \;=\; V_{\rm eff}(\varphi_{\rm min}) \;=\; V_{\rm min}^{(0)}
	\,+\, g_3^2 V_{\rm min}^{(1)} \,+\, g_3^4 V_{\rm min}^{(2)}
	\,+\, {\cal O}(g_3^6) \quad.
\ee
We have to distinguish two cases

(i) One tree-level extremum is at
$\varphi_{\rm min}^{(0)} = 0$ and will stay there
($\varphi_{\rm min}^{(1)} = \varphi_{\rm min}^{(2)} = 0$).
{}From equation (\ref{Vmin}) on gets:
\bea
V_{\rm min}^{(0)} &=& 0 \quad,\\
V_{\rm min}^{(1)} &=& -\frac{1}{3\pi} (-\lambda v_0^2)^{3/2}\quad,\\
V_{\rm min}^{(2)} &=& \frac{3}{64\pi^2}
	\lambda v_0^2
	\left(-3 - 8 \lambda +
	4 \ln\left( 2\sqrt{-\lambda v_0^2} \right) \right) \quad.
\eea
$V_{\rm min}^{(1)}$ and $V_{\rm min}^{(2)}$ are real for $v_0^2 < 0$,
i.e.\ above the tree-level roll-over temperature, and complex below.
This is due to the fact that $\varphi_{\rm min}^{(0)} = 0$ is a maximum of
$V_{\rm ht}(\varphi)$ for $v_0^2 > 0$.

(ii) At this temperatures one should expand around the broken minimum
$\varphi_{\rm min}^{(0)} = v_0$
\bea
\varphi_{\rm min}^{(1)} &=&
\frac{3}{32 \lambda \pi}
\left( 1 + 2\sqrt{\xi}\lambda + 2^{5/2}\lambda^{3/2} \right) \quad, \\
V_{\rm min}^{(0)} &=& -\frac{\lambda v_0^4}{4} \quad,\label{Vmin0}\\
V_{\rm min}^{(1)} &=& -\frac{1}{48\pi}
	\left(3 + 2^{7/2}\lambda^{3/2} \right)\, v_0^3 \quad,\\
V_{\rm min}^{(2)} &=&
	 \frac{3}{1024 \pi^2 \lambda} \;v_0^2\;
	\bigg( \! -3 +
	(11 + 42 \ln(2/3)) \lambda
	- 2^{9/2} \lambda^{3/2}
  	+ 24\lambda^2 - 2^{11/2}\lambda^{5/2}
  	- 128\lambda^3 \nonumber\\
	&&+\; 8 \lambda \left(1 - 4 \lambda + 8 \lambda^2 \right)
		\ln\left( 1 + \sqrt{2\lambda} \right)
	+ 64 \lambda^3 \ln\left( 3 \sqrt{2\lambda} \right)\nonumber\\
	&&
	+\; 2 \lambda ( -17 - 16 \lambda + 64 \lambda^2)
		\ln\left(\frac{v_0}{\mu}\right)
	\bigg) \quad.\label{Vmin2}
\eea
The values of  $V_{\rm min}^{(i)}$ ($i=1,2,3$) are independent of $\xi$
as they should. Note that we do not have any IR-divergences showing up
in covariant non-background gauges \cite{Laine}.

Equations (\ref{Vmin0}) - (\ref{Vmin2}) show that the effective
expansion parameter at small $\lambda$ is
{\Large $\frac{g_3^2}{4\pi\lambda}$}. Consequently it
becomes of order 1 if one approaches the critical temperature from below.
Going to the limit $\lambda \rightarrow$ {\em small}  therefore does not help
in
improving the convergence of the loop expansion close to the critical
temperature.  The overall $\frac{1}{\lambda}$ in Eq. (\ref{Vmin2}) arises
entirely from inserting $\varphi_{\rm min}^{(1)}$ into the tree- and one-loop
potential.

\begin{figure}[t]
\begin{picture}(14.0,8.5)
\put(0.7,6.8){$g_3^2$}
\put(5,6.5){$\xi=0$}
\put(5,6.05){$\xi=1$}
\put(5,5.6){$\xi=2$}
\put(1.5,0.3){\epsfxsize12cm \epsffile{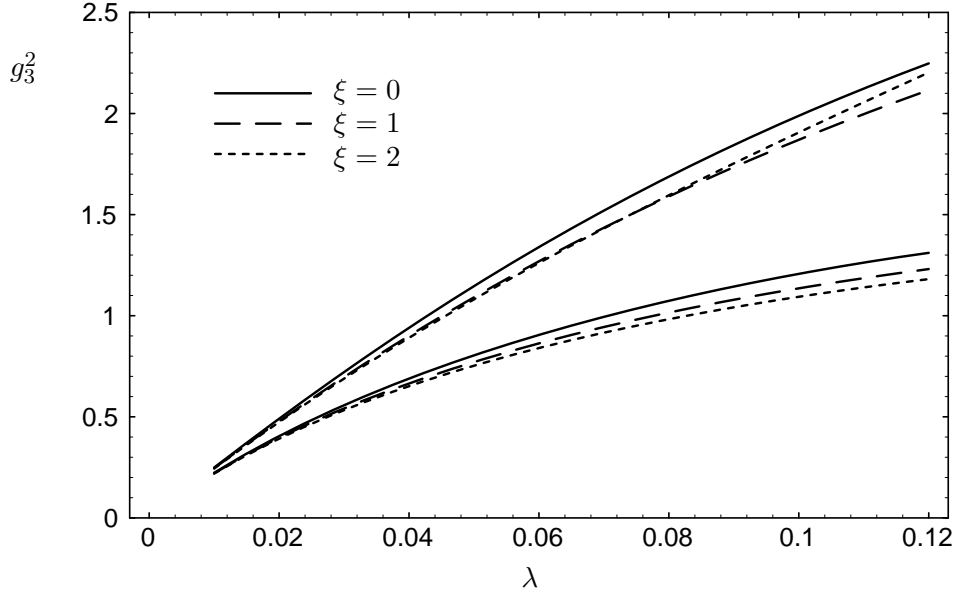}}
\put(7.5,0){$\lambda$}
\end{picture}
\caption{$g_3^2$ vs.~$\lambda$ at the critical temperature}
\label{FIGg3sla}
\end{figure}

Although the expansion around $\varphi_{\rm min}^{(0)} = v_0$
is manifestly gauge-fixing independent, it is not
useful close to the critical temperature. As soon as
$\varphi_{\rm min}^{(0)}$ tends
towards zero the true position of the minimum may no longer be considered
as being obtained as a small perturbation around $v_0(T)$. In the following,
we therefore do no longer insist on the $g_3^2$ expansion of
$\varphi_{\rm min}$.
Consequently, we do no longer work consistently to a given order in $g_3^2$,
and some $\xi$ dependence must be expected. A small $\xi$ dependence
would be an indication of a reasonable convergence of the approximation.

Both the one loop and the two-loop potential predict a first order phase
transition. They have two local minima, which are degenerated at the
respective critical temperature.
Up to now the value of $v$ used to rescale the field in equation
(\ref{rescale}), is
arbitrary. It is an appropriate choice to take $v$ and $\mu$
to be the value of the
scalar field at the broken minimum. The asymmetric minimum of the rescaled
field $\varphi$ is then at $\varphi_a = 1$
\begin{equation}\label{brokenmin}
V_{\rm eff}'(\varphi\!=\! 1) \;=\; 0 \quad.
\end{equation}
In addition, we have  the condition
\begin{equation}\label{criticaltem}
V_{\rm eff}(\varphi\!=\!0) \;=\; V_{\rm eff}(\varphi\!=\!1) \quad.
\end{equation}
at the  critical temperature.
$g_3^2$ and $v_0^2$, the two parameters of the high temperature effective
action (\ref{Sht}), can be calculated from equations (\ref{brokenmin}) and
(\ref{criticaltem}).
{}From this it follows that
the two coupling constants $\lambda$ and $g_3^2$ are not
independent at the critical temperature. Since $g_3^2$ determines the
sphaleron rate this relation is important to determine cosmological
bounds on the Higgs mass \cite{HellmundEA}.

In Figure \ref{FIGg3sla}, $g_3^2$ is taken at the corresponding one
and two-loop critical temperature, respectively, and is plotted versus
$\lambda$ for $\xi=0,1,2$. The gauge-fixing dependence is weak,
both at one and two-loop.
The inclusion of the two-loop contributions changes the magnitude of
$g_3(T_c)^2$ by about 40\%. This would reduce the sphaleron rate by many
orders of magnitude. Therefore bounds on the Higgs mass from the wash-out
of the baryon asymmetry may be less reliably than thought so far.

\begin{figure}[t]
\begin{picture}(14.0,8.5)
\put(0.5,6.5){\Large $\frac{\varphi_{\rm min}(T)}{T}$}
\put(2,0.2){\epsfxsize12cm \epsffile{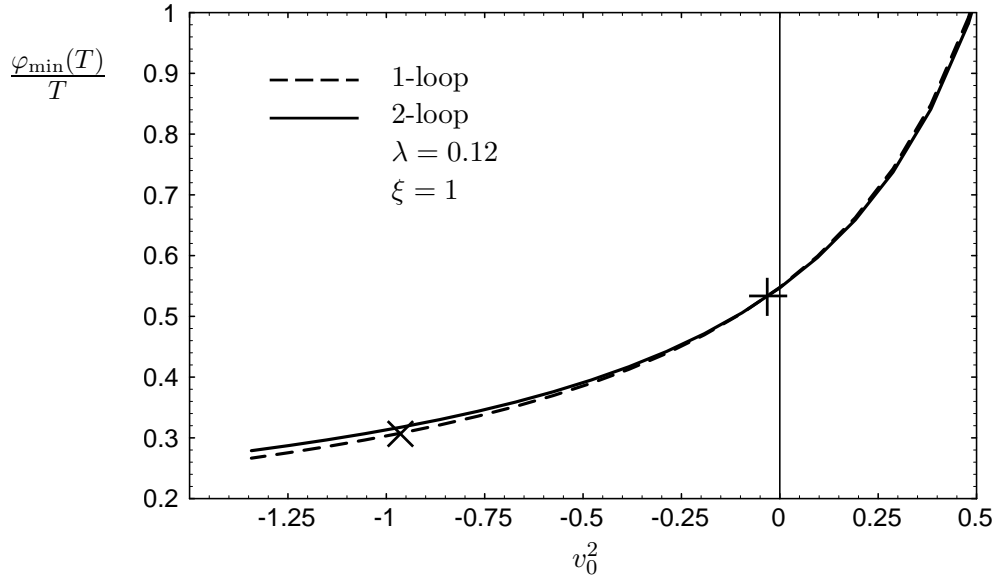}}
\put(5.6,6.5){1-loop}
\put(5.6,6){2-loop}
\put(5.6,5.5){$\lambda = 0.12$}
\put(5.6,5){$\xi=1$}
\put(8,0.1){$v_0^2$}
\end{picture}
\caption{The field value at the broken minimum in units of the temperature
vs.~$v_0^2$ for $\lambda = 0.12$ and $\xi=1$}
\label{FIGvTv0}
\end{figure}

The large corrections to $g_3(T_c)^2$ are not caused by large corrections
to $\varphi_{\rm min}(T)\,/\,T$ at fixed temperature, but
by the shift of $T_c$. This is demonstrated in figure \ref{FIGvTv0}.
(The $\xi$-dependence is again not significant her.) Crosses denote
the one ($\times$) and two (+) loop critical temperatures. Going from one
to two-loop one essentially moves along an almost universal curve.

The determination of the critical temperature from the perturbative
potential is of course very questionable because the latter is unreliable
for small $\varphi$ values. Still, it is remarkable that the
two-loop potential leads to a lowering of $T_c$ and an increase
of $g_3^2$ almost independently of the gauge-fixing parameter $\xi$.

\begin{figure}[t]
\begin{picture}(14.0,8.5)
\put(0.5,6.5){\Large $\frac{V_{\rm eff}^{(2)}}{g^2\, v^4}$}
\put(0.5,4){\Large $\frac{V_{\rm eff}^{(1)}}{g^2\, v^4}$}
\put(1.5,0.2){\epsfxsize12cm \epsffile{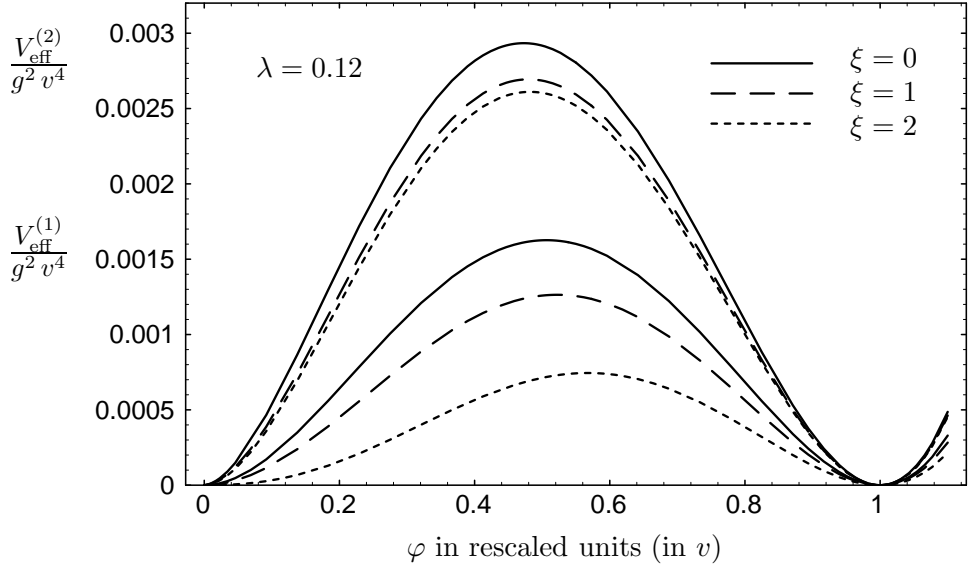}}
\put(3.8,6.5){\parbox{3cm}{$\lambda = 0.12$}}
\put(11.7,6.6){$\xi=0$}
\put(11.7,6.15){$\xi=1$}
\put(11.7,5.7){$\xi=2$}
\put(5.8,0.1){$\varphi$ in rescaled units (in $v$)}
\end{picture}
\caption{The one and two-loop effective potential for $\lambda = 0.12$
and different values of the gauge-fixing parameter $\xi$}
\label{FIGVeffTc}
\end{figure}

The one and two-loop potential at the corresponding critical temperatures
is plotted for $\xi=0,1,2$ in figure \ref{FIGVeffTc}.
One first notices that the bulge between the
symmetric and the asymmetric minimum grows from one to two loop, which
is in agreement with previous calculations using
Landau gauge \cite{ArnoldEs,FodorHe}.
Accordingly, the phase transition is stronger first order, as predicted
by lattice calculations \cite{BunkEA, FarakosEA2}.
While $V_{\rm eff}^{(1)}(\varphi)$ depends strongly on
$\xi$, the shape of $V_{\rm eff}^{(2)}(\varphi)$ is only slightly
$\xi$-dependent.
This indicates that the loop-expansion might asymptotically converge towards
an $\xi$-in\-de\-pen\-dent effective potential if one uses 't~Hooft background
gauges.

$v(T)$, the field value of the broken minimum used for rescaling
(equations (\ref{rescale}, \ref{brokenmin})), is not a physical
observable. It turns out to be $\xi$-dependent.
We have calculated the $Z$-factor for the Higgs-kinetic term at one loop
\begin{eqnarray}
Z_H(\varphi) &=& 1 \;+\;  \frac{g_3^2}{4 \pi} \; \Bigg\{
  -\; \frac{3}{m_{\rm Gs}+m_W} \;
   +\;  \frac{1}{m_W^2} \left(
		\frac{m_{\rm Gs}^3-m_W^3}{m_{\rm Gs}^2-m_W^2} \;- \;
		\frac{m_{\rm Gs}^3-\xi^{3/2}m_W^3}{m_{\rm Gs}^2-\xi m_W^2}
	\right) \nonumber\\
 &&\qquad\qquad  +\;\frac{10 - 13\sqrt{\xi} + 9\xi}{16\,(1 + \sqrt{\xi})}
\; \frac{1}{m_W^3}
   \left(\frac{\partial m_W^2}{\partial \varphi} \right)^2
 \;-\; \frac{2}{16}\, \frac{1}{m_{\rm gh}^3}
      \left(\frac{\partial m_{\rm gh}^2}{\partial \varphi} \right)^2
 \nonumber\\
&&   \qquad\qquad +\; \frac{1}{16}\, \frac{1}{m_{\rm Gs}^3}
      \left(\frac{\partial m_{\rm Gs}^2}{\partial \varphi} \right)^2
    \;+\; \frac{1}{48}\, \frac{1}{m_{\rm H}^3}
      \left(\frac{\partial m_{\rm H}^2}{\partial \varphi} \right)^2
\quad\Bigg\}
\end{eqnarray}
It has a strong $\xi$-dependence for small $\varphi$-values and becomes
even negative at some range as already pointed out in
reference \cite{KripfganzEA}. Near the broken minimum however it behaves
well but is still $\xi$-dependent.
One can discuss\footnote{ We thank C.~Wetterich for raising this question.},
if this $\xi$-dependence cancels the one of $v(T)$ calculating
the renormalized field value
$\sqrt{Z(v(T))} v(T)$. We found that there is
indeed some 45\% reduction if the one-loop $Z$-factor with
the two-loop $g_3^2$ and the two-loop $v(T)$ is used.
Calculating the temperature dependent $W$-mass would also require
the $Z_W$-factor.

In conclusion, we have demonstrated the applicability of the
class of 't~Hooft background gauges (eq.~\ref{gauge-fixing}) for studies of
the electroweak phase transition.
The main advantage of this class of gauge-fixings is the
absence of IR-divergences in the broken phase, which are caused by
massless Goldstone bosons in the class of covariant gauges
(${\cal F}^a = \partial_i A_i^a$) which is prominent in literature.
The latter class has been used by M.~Laine in a recent publication
\cite{Laine} to calculate the two loop potential. He found that the loop
expansion converges even in the broken phase only for small values of $\xi$.
This is essentially due to a $\xi$-dependent infrared divergence which does
not show up in 't~Hooft background gauges. In our opinion this shows the
superiority of these gauges.
The severe problems of perturbation theory in the symmetric phase
caused by nonperturbative condensates can of course not be
cured either.

Besides the improved IR-behavior in the broken phase there are some technical
advantages due to the absence of a mixed Goldstone-gauge boson propagator.
If one restricts oneself to the 't~Hooft-Feynman gauge ($\xi=1$), the
gauge boson propagator turns out to be quite simple as well
(cf.~eq.~(\ref{gauge_prop})).
Note that the background gauges are also used in the computation
of  high energy cross
sections\cite{DennerEA}.

We showed explicitly that the value of the effective potential at
its minima is independent of the gauge-fixing parameter order by order
if it is expanded consistently in $g_3^2$.

Close to the critical temperature this expansion breaks down.
Here we worked with the full two-loop potential and used a rescaling
procedure which is especially suited to the treatment of
quasiclassical solutions like critical bubbles \cite{KripfganzEA}
and sphalerons \cite{HellmundEA}.
This procedure is not gauge-fixing independent but the $\xi$-dependence
becomes substantially weaker from one to two-loop order
 in the case of the effective potential
(figure \ref{FIGVeffTc}).

\newpage

{\Large\bf\parindent0cm Appendix} \\[2ex]
The propagators and vertices are read off equation (\ref{Sfluct}).
We used the form of the gauge boson propagator given in
equation (\ref{gauge_prop}). It is straightforward to write down
the two-loop graphs shown in figure \ref{FIGgraphs}.
While the ``figure 8'' graphs can be evaluated easily the momenta
in the numerators of the ``sunset'' graphs have to be removed first.

We did this in two steps:\\
(i) after use of momentum conservation two loop-momenta $k$
and $p$ are left. The mixed scalar products $kp$ are removed
using the following identities:
\bea
\frac{2 kp}{(k+p)^2 + m^2} &=&
1 \;-\; \frac{k^2 + p^2 + m^2}{(k+p)^2 + m^2}\\[1ex]
\int \frac{d^dp}{(2 \pi)^d}\;
\;kp\; F(p^2, k^2)
&= & 0 \\[1ex]
\int \frac{d^dp}{(2 \pi)^d}\;
(kp)^2 F(p^2, k^2)
& =&
\frac{1}{d} \int \frac{d^dp}{(2 \pi)^d}\;
k^2 p^2 F(p^2, k^2)
\eea
(ii) The remaining momenta in the numerators are removed using the identities:
\bea
\frac{k^2}{k^2 + m^2} \quad=\quad
1 \;-\; \frac{m^2}{k^2 + m^2}\\[1ex]
\int \frac{d^dk}{(2 \pi)^d}\; \label{Profiformel}
\int \frac{d^dp}{(2 \pi)^d}\quad
\frac{k^m p^n}{k^2 + m^2}
 \quad=\quad 0
\eea
Formula (\ref{Profiformel}) holds only due to the cancelation of
IR-singularities.
This reduction procedure has been performed with FORM.
The integrals left are of one of the types given in equations
(\ref{IntG}, \ref{IntH}).

The single graphs are given below where the combinatorical factors
are chosen in a way that the contribution to the two-loop potential
is given by:
\be
g_3^4 \;\sum{\rm figure~8's}
\quad-\quad \frac{1}{2}\, g_3^4\; \sum{\rm sunsets}
\ee
The dimension is $d=3-2\epsilon$.
\begin{eqnarray*}
&&\!\!\!\!\!   {\rm WWW} = \\
&&       H(\mW ,\mgh ,\mgh ) \left(
          - 3\, \xi\, \mWq
          + \frac{3}{4}\, \mWq
          \right)
      \;+\; H(\mW ,\mW ,\mW ) \left(
          - 9\, \mWq \, d
          + \frac{45}{4}\, \mWq
          \right)\\
&&       + G(\mgh ,\mgh ) \left( \!
          - \frac{3}{4}
          + \frac{3}{2}\, \xi
          - 3 \xi^2\, \frac{1}{d}
          + 3 \xi^2
          \right)
       \,+\, G(\mW ,\mgh ) \left(
          \frac{3}{2}
          + 6\, \xi\, \frac{1}{d}
          + 6\, \xi\, d
          - 12\, \xi
          \right)\\
&&       +\; G(\mW ,\mW ) \left(
          - \frac{9}{4}
          - 3\, \frac{1}{d}
          + 3\, d
          \right)\\[2ex]
&&\!\!\!\!\!   {\rm WGhGh} = \\
&&      H(\mgh ,\mgh ,\mgh ) \left(
          - \frac{3}{2}\, \xi^2\, \mWq
          \right)
     \;+\; H(\mW ,\mgh ,\mgh ) \left(
          + 6\, \xi\, \mWq
          - \frac{3}{2}\, \mWq
          \right)\\
&&     +\; G(\mgh ,\mgh ) \left(
          + \frac{3}{2}
          - \frac{3}{2}\, \xi
          \right)
     \;+\; G(\mW ,\mgh ) \left(
          - 3
          \right)\\[8ex]  
&&\!\!\!\!\!   {\rm WGsGs} = \\
&&        H(\mW ,\mGs ,\mGs ) \left(
          + \frac{3}{4}\, \mWq
          - 3\, \mGsq
          \right)
       \;+\; G(\mW ,\mGs ) \left(
          + \frac{3}{2}
          \right)\\
&&       +\; G(\mGs ,\mgh ) \left(
          + \frac{3}{2}\, \xi
          \right)
       \;+\; G(\mGs ,\mGs ) \left(
          - \frac{3}{4}
          \right)\\[2ex]
&&\!\!\!\!\!   {\rm GhGhGs} = \\
&&       + H(\mGs ,\mgh ,\mgh ) \left(
          + \frac{3}{8}\, \xi^2\, \varphi^2
          \right)\\[2ex]
&&\!\!\!\!\!   {\rm WW} = \\
&&        G(\mgh ,\mgh ) \left(
          - \frac{3}{2}\, \xi^2\, \frac{1}{d}
          + \frac{3}{2}\, \xi^2
          \right)
       \;+\; G(\mW ,\mgh ) \left(
          + 3\, \xi\, \frac{1}{d}
          + 3\, \xi\, d
          - 6\, \xi
          \right)\\
&&       +\; G(\mW ,\mW ) \left(
          + \frac{9}{2}
          - \frac{3}{2}\, \frac{1}{d}
          - \frac{9}{2}\, d
          + \frac{3}{2}\, d^2
          \right)\\[2ex]
&&\!\!\!\!\!   {\rm WGs} = \\
&&       G(\mW ,\mGs ) \left(
          - \frac{9}{8}
          + \frac{9}{8}\, d
          \right)
       \;+\; G(\mGs ,\mgh ) \left(
          + \frac{9}{8}\, \xi
          \right)\\[2ex]
&&\!\!\!\!\!   {\rm GsGs} = \\
&&       + G(\mGs ,\mGs ) \left(
          + \frac{15}{4}\, \lambda
          \right)\\[2ex]
&&\!\!\!\!\!   {\rm WGsH} = \\
&&      H(\mW ,\mGs ,\mH ) \left(
          - \frac{3}{2}\, \frac{\mGsq \, \mHq}{\mWq }\,
          + \frac{3}{4}\, \frac{ \mGsf}{\mWq }\,
          + \frac{3}{4}\, \frac{ \mHf}{\mWq }\,
	  + \frac{3}{4}\, \mWq
          - \frac{3}{2}\, \mGsq
          - \frac{3}{2}\, \mHq
          \right)\\
&&     +\; H(\mGs ,\mH ,\mgh ) \left(
          + \frac{3}{2}\, \frac{1}{\mWq }\, \mGsq \, \mHq
          - \frac{3}{4}\, \frac{1}{\mWq }\, \mGsf
          - \frac{3}{4}\, \frac{1}{\mWq }\, \mHf
          \right)\\
&&       +\; G(\mW ,\mGs ) \left(
          + \frac{3}{4}
          + \frac{3}{4}\, \frac{1}{\mWq }\, \mGsq
          - \frac{3}{4}\, \frac{1}{\mWq }\, \mHq
          \right)\\
&&       +\; G(\mW ,\mH ) \left(
          + \frac{3}{4}
          - \frac{3}{4}\, \frac{1}{\mWq }\, \mGsq
          + \frac{3}{4}\, \frac{1}{\mWq }\, \mHq
          \right)\\
&&       +\; G(\mGs ,\mgh ) \left(
          + \frac{3}{4}\, \xi
          - \frac{3}{4}\, \frac{1}{\mWq }\, \mGsq
          + \frac{3}{4}\, \frac{1}{\mWq }\, \mHq
          \right)
       \;+\; G(\mGs ,\mH ) \left(
          - \frac{3}{4}
          \right)\\
&&       +\; G(\mH ,\mgh ) \left(
          + \frac{3}{4}\, \xi
          + \frac{3}{4}\, \frac{1}{\mWq }\, \mGsq
          - \frac{3}{4}\, \frac{1}{\mWq }\, \mHq
          \right)\\[2ex]
&&\!\!\!\!\!   {\rm GsGsH} = \\
&&       H(\mGs ,\mGs ,\mH ) \left(
          + 6\, \lambda^2\, \varphi^2
          \right)\\[2ex]
&&\!\!\!\!\!   {\rm GhGhH} = \\
&&       H(\mH ,\mgh ,\mgh ) \left(
          - \frac{3}{16}\, \xi^2\, \varphi^2
          \right)\\[8ex]  
&&\!\!\!\!\!   {\rm WWH} = \\
&&       H(\mW ,\mW ,\mH ) \left(
          + \frac{3}{32}\, \varphi^2\, \frac{1}{\mWf }\, \mHf
          - \frac{3}{8}\, \varphi^2\, \frac{1}{\mWq }\, \mHq
          + \frac{3}{8}\, \varphi^2\, d
          - \frac{3}{8}\, \varphi^2
          \right)\\
&&     +\; H(\mW ,\mH ,\mgh ) \left(
         \frac{3}{8} \xi \varphi^2 \frac{\mHq}{\mWq }
          + \frac{3}{8} \xi \varphi^2
          - \frac{3}{16} \xi^2 \varphi^2
          - \frac{3}{16} \varphi^2 \frac{\mHf}{\mWf }
          + \frac{3}{8} \varphi^2 \frac{\mHq}{\mWq }
          - \frac{3}{16} \varphi^2
          \right)\\
&&     +\; H(\mH ,\mgh ,\mgh ) \left(
          - \frac{3}{8}\, \xi\, \varphi^2\, \frac{1}{\mWq }\, \mHq
          + \frac{3}{8}\, \xi^2\, \varphi^2
          + \frac{3}{32}\, \varphi^2\, \frac{1}{\mWf }\, \mHf
          \right)\\
&&     +\; G(\mgh ,\mgh ) \left(
          + \frac{3}{16}\, \xi\, \varphi^2\, \frac{1}{\mWq }
          - \frac{3}{32}\, \varphi^2\, \frac{1}{\mWf }\, \mHq
          \right)\\
&&     +\; G(\mW ,\mgh ) \left(
          - \frac{3}{16}\, \xi\, \varphi^2\, \frac{1}{\mWq }
          + \frac{3}{16}\, \varphi^2\, \frac{1}{\mWf }\, \mHq
          - \frac{3}{16}\, \varphi^2\, \frac{1}{\mWq }
          \right)\\
&&     +\; G(\mW ,\mW ) \left(
          - \frac{3}{32}\, \varphi^2\, \frac{1}{\mWf }\, \mHq
          + \frac{3}{16}\, \varphi^2\, \frac{1}{\mWq }
          \right)\\
&&     +\; G(\mW ,\mH ) \left(
          + \frac{3}{16}\, \xi\, \varphi^2\, \frac{1}{\mWq }
          - \frac{3}{16}\, \varphi^2\, \frac{1}{\mWq }
          \right)\\
&&     +\; G(\mH ,\mgh ) \left(
          - \frac{3}{16}\, \xi\, \varphi^2\, \frac{1}{\mWq }
          + \frac{3}{16}\, \varphi^2\, \frac{1}{\mWq }
          \right)\\[2ex]
&&\!\!\!\!\!   {\rm HHH} = \\
&&       H(\mH ,\mH ,\mH ) \left(
          + 6\, \lambda^2\, \varphi^2
          \right)\\[2ex]
&&\!\!\!\!\!   {\rm WH} = \\
&&       G(\mW ,\mH ) \left(
          - \frac{3}{8}
          + \frac{3}{8}\, d
          \right)
     \;+\; G(\mH ,\mgh ) \left(
          + \frac{3}{8}\, \xi
          \right)\\[2ex]
&&\!\!\!\!\!   {\rm GsH} = \\
&&       G(\mGs ,\mH ) \left(
          + \frac{3}{2}\, \lambda
          \right)\\[2ex]
&&\!\!\!\!\!   {\rm HH} = \\
&&       + G(\mH ,\mH ) \left(
          + \frac{3}{4}\, \lambda
          \right)\\[2ex]
&&\!\!\!\!\!   {\rm GhGs} = \\
&&       G(\mGs ,\mgh ) \left(
          + \frac{3}{4}\, \xi
          \right)
\end{eqnarray*}

\end{document}